\documentclass[12pt]{article}
\usepackage{latexsym}                 
\usepackage[T1]{fontenc}
\usepackage{lmodern}
\usepackage[latin1]{inputenc}
\usepackage{latexsym}                 
\usepackage{color}                 
\usepackage{amssymb}
\usepackage{amsmath}
\usepackage{amsthm}
\usepackage{graphicx}
\usepackage{hyperref}
\usepackage{verbatim}
\usepackage{mathptmx}      
\newcommand{\eq}{\end{equation}}
\def\b1{\mathbf 1}

\def\biglf{\par\bigskip\noindent}
\def\eref#1{(\ref{#1}%
)}

\def\RSlabel#1{\label{#1}%
}

\newcommand{\bql}[1]{%
\begin{equation}\label{#1}%
}

\def\filename#1{}
\begin{document}
\begin{center}
 
  {\Large\bf Modelling Recovered Cases and Death Probabilities\\~\\
    for the COVID-19 Outbreak} 
  \\~\\
  Robert Schaback\footnote{Institut f\"ur Numerische und Angewandte Mathematik,
Universit\"at G\"ottingen,
Lotzestra\ss{}e 16-18,
D-37073 G\"ottingen,
Germany, schaback@math.uni-goettingen.de}\\~\\
  March 26th, 2020
\end{center} 

{\bf Abstract}: From March 23rd, the  data for the recovered cases of
COVID-19 are missing from the standard repository maintained by the Johns
Hopkins University
in collaboration with the WHO. But since data concerning recovered patients are
extremely important for modelling the COVID-19 outbreak,
a method for estimating the missing data is provided and tested.
As a byproduct, it produces estimates for the probabilities
to die $k$ days after confirmation, or to survive after $d$ days.

\section{Available Data}\RSlabel{SecAD}
The COVID-19 repository
\cite{repository} at Johns Hopkins University provides cumulative data on
\begin{itemize}
\item {\em Confirmed} infected persons ($C$) and
\item {\em Deaths} ($D$)
\end{itemize} 
per territory, as integer-valued time series beginning on Jan. 22nd, 2020.
Up to March 23rd, there was a count of  {\em Recovered} ($R$)
as well,
but this was stopped for various reasons. One reason that applies for Germany
is that there simply is no defined registration for the Recovered.
Authorities only report confirmed cases and deaths, but they say
\cite{rkiModellierung}
that a person with mild or no symptoms is considered to be
recovered after 14 days. 
\biglf
However, the Recovered are extremely important for
analysis of the COVID-19 outbreak. They are the major $R$ part of
standard models like SIR and SEIR \cite{hethcote:2000-1},
and they define the
{\em Infectious} $I=C-D-R$ that hopefully have a peak at some time from
the exponential outbreak. Prediction of this peak is the central problem
for any analysis of the COVID-19 pandemic.  
\section{A First Failure}\RSlabel{SecFF}
To illustrate the problem, let us assume a form of the 14-day-rule:
\begin{itemize}
\item[] If a confirmed person survives for $d$ days after confirmation,
 the person is counted as recovered.  
\end{itemize} 
To analyze this, let $D_{i,j}$ be the count
of deaths on day $j$ of the new registered of day $i$, i.e. $i<j\leq i+d$. Then
the new casualties on day $j$ are 
$$
D_j-D_{j-1}=\sum_{i=j-d}^{j-1}D_{i,j}
$$
and the new Recovered $R_j$ at day $j$ are the new Confirmed at day $j-d-1$
minus all casualties of Confirmed at day $j-d-1$ and died
up to day $j-1$, i.e.
$$
R_j-R_{j-1}=C_{j-d-1}-C_{j-d-2}-\sum_{k=j-d}^{j-1}D_{j-d-1,k}.
$$
Summing over $j$ up to some $n$ yields
$$
\begin{array}{rcl}
  R_n
  &=&
  C_{n-d-1}-\displaystyle{\sum_{j=-\infty}^n\sum_{k=j-d}^{j-1}D_{j-d-1,k}}\\
  &=&
  C_{n-d-1}-\displaystyle{\sum_{k=-\infty}^{n-1}\sum_{i=k-d}^{k-1}D_{i,k}}\\
  &=&
  C_{n-d-1}-\displaystyle{\sum_{k=-\infty}^{n-1}(D_k-D_{k-1})}\\
  &=&
  C_{n-d-1}-D_{n-1}
\end{array}
$$
which is quite a simple formula, but the count should be cumulative.
This fails for various countries, unless $d$ is chosen to be
intolerably small. For example, the data for Italy
at day $n=52$ (March 14th) 
have $C_{n-14}-D_{n}=888-1266=-378$.
Furthermore, while the general assumption makes sense for large $d$,
the above solution will then necessarily fail.
\section{Probabilistic Approach}\RSlabel{SecPA}
Assume that $p_k$ is the probability to die
at day $k$ after confirmation, after still alive at day $k-1$.
These events are assumed to be stochastically independent.
The probability to die at day $k$ after confirmation then is
\bql{qkpk}
q_k=p_k\prod_{j=1}^{k-1} (1-p_j).
\eq
Now the deaths at day $n$ can be expected to be
\bql{maineq}
D_n-D_{n-1}=\sum_{m=1}^dq_m(C_{n-m}-C_{n-1-m}),
\eq
and it is clear how to set up a nonlinear constrained least-squares fit
to estimate the probabilities. The above equations are set up for various $n$,
using the newest available data, and leading to more than $d$ equations
for the $d$ unknowns $p_1,\ldots,p_d$ that enter nonlinearly
into the $q_k$ by \eref{qkpk}. Then the sum of squares of the residuals is
minimized. 
\biglf
The probability to be still alive on day $d$ after confirmation
is
$$
\prod_{j=1}^{d}(1-p_j)
$$
and therefore the Recovered can be modelled by 
$$
R_n-R_{n-1}=(C_{n-d}-C_{n-d-1})\prod_{j=1}^{d}(1-p_j).
$$
This is automatically cumulative and maintains the integrity of
person counts between the Confirmed, Deaths, and Recovered.
All confirmed persons end up some time later as recovered or dead.
\biglf
As a byproduct, the above approach provides estimates for the probability
to die on day $k$ after confirmation,
or to survive after $d$ days. Of course, this is strongly dependent on how
and when confirmations are done in a certain territory,
and it only applies to persons with confirmed COVID-19
by the rules applied there. 
\section{Examples}\RSlabel{SecEx}
Table \ref{table1} shows death and survival probabilities for
the Confirmed of various territories, using data available
on day  64, March 26th. These results were obtained for
a fixed decision time of $d=14$ days,
and setting up $2d$ equations \eref{maineq} for the latest available data.
The results will vary somewhat
if $d$ is changed or if more data are used, but we wanted to avoid any
country-dependent tuning.
\biglf
The corresponding figures show the Confirmed ($C$) and Deaths ($D$)
from the Johns Hopkins data source after smoothing with the weights
$(1,4,6,4,1)/16$. The $R$ values are the  Recovered estimated by the above
procedure,
entering into the Infectious $I=C-R-D$. The dotted green values are the $R$
data from the old data set, as far as available. 
The deviations from the dotted and the continuous $R$ values
are not necessarily failures of the method, because the old $R$ data
were questionable anyway and might not maintain data consistency.
France has not reported any Recovered for quite some
time, and the data from Germany is doubtful for the
procedural reasons mentioned above. 
\biglf
The survival probabilities in the last row of Table \ref{table1}
are low for countries before and after the full outbreak, while
they are large for countries in a critical phase.
When testing different decision times $d$, in particular larger $d$,
the
estimated survival probabilities do not
change much, while the death probabilities
continue their
tendency to have peaks around day 5 and day 14. There is no  attempt
to interpret the findings to more detail, in particular towards
medical, social, or political implications. In further work,
the obtained data for the Recovered should enter into
predictions of a possible peak of the Infectious. 
\begin{table}
  \begin{center}
\begin{tabular}{||r||l|l|l|l|l|l||}
  \hline\hline
  Days &France & Germany & Italy & Spain & US & Hubei\\
\hline\hline
1 & 0 & 0.001 & 0.073 & 0.016 & 0.01 & 0\\
2 & 0 & 0 & 0 & 0.014 & 0 & 0\\
3 & 0 & 0 & 0 & 0.018 & 0 & 0.019\\
4 & 0 & 0 & 0.003 & 0 & 0 & 0\\
5& 0.031 & 0.001 & 0.013 & 0 & 0.002 & 0\\
6& 0.011 & 0.007 & 0.025 & 0 & 0 & 0.022\\
7& 0 & 0 & 0 & 0 & 0 & 0.047\\
8&0 & 0 & 0 & 0.034 & 0 & 0\\
9&0 & 0 & 0 & 0.009 & 0 & 0\\
10& 0 & 0 & 0 & 0.164 & 0 & 0\\
11& 0 & 0 & 0 & 0.053 & 0 & 0\\
12& 0 & 0.002 & 0 & 0 & 0.026 & 0\\
13 &0 & 0 & 0 & 0 & 0.155 & 0\\
14& 0.394 & 0.018 & 0.085 & 0 & 0 & 0\\
\hline
15& 0.581 & 0.97 & 0.814 & 0.722 & 0.813 & 0.915\\
\hline\hline
\end{tabular}  
  \end{center}
  \caption{Probabilities for confirmed persons to die at
    days 1 to 14 after confirmation, and to survive to
    day 15. Based on a test run on day 64 (March 26th).\label{table1}}
\end{table}
\begin{figure}
\begin{center}
\includegraphics[width=6.0cm,height=5.0cm]{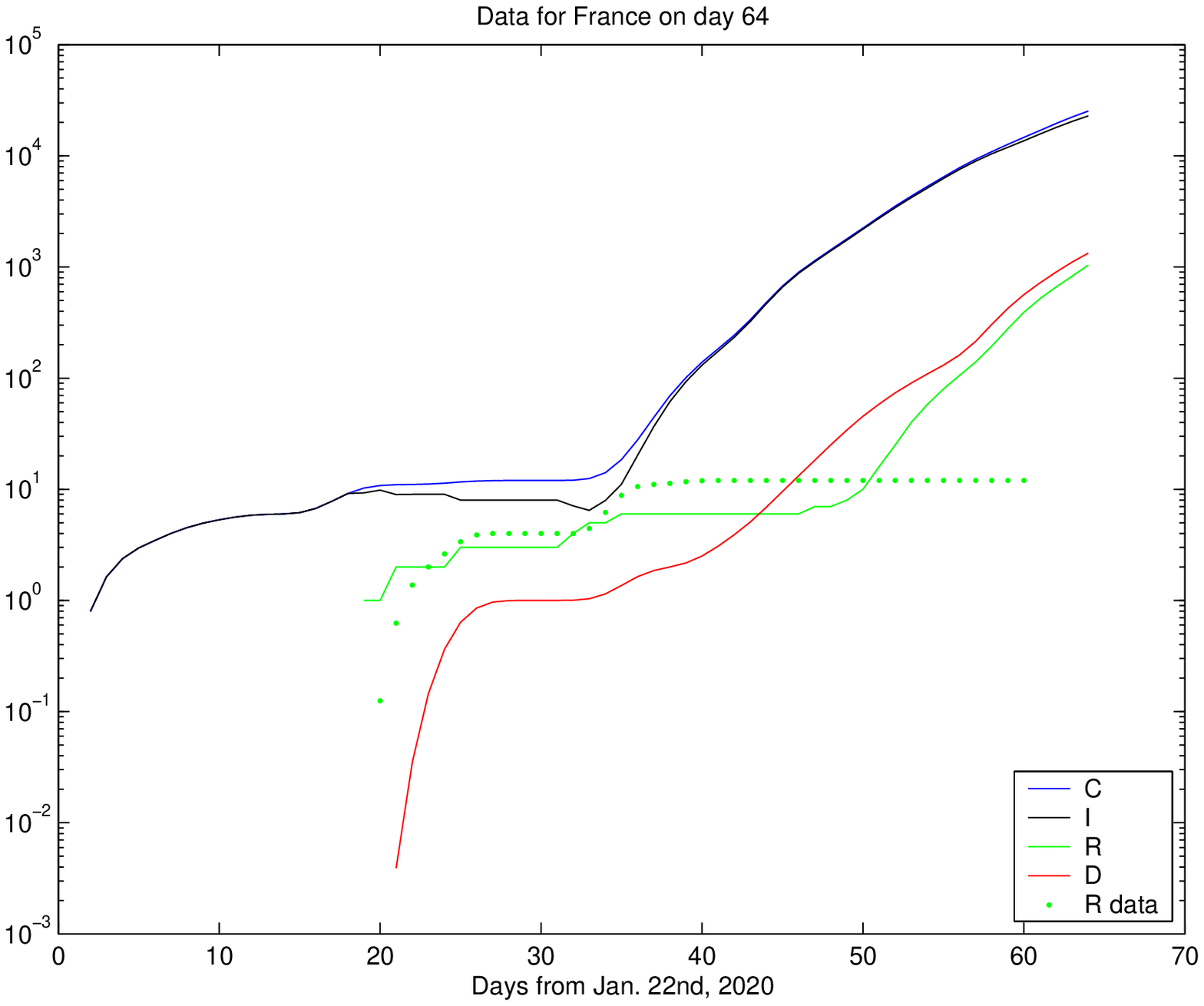}
\includegraphics[width=6.0cm,height=5.0cm]{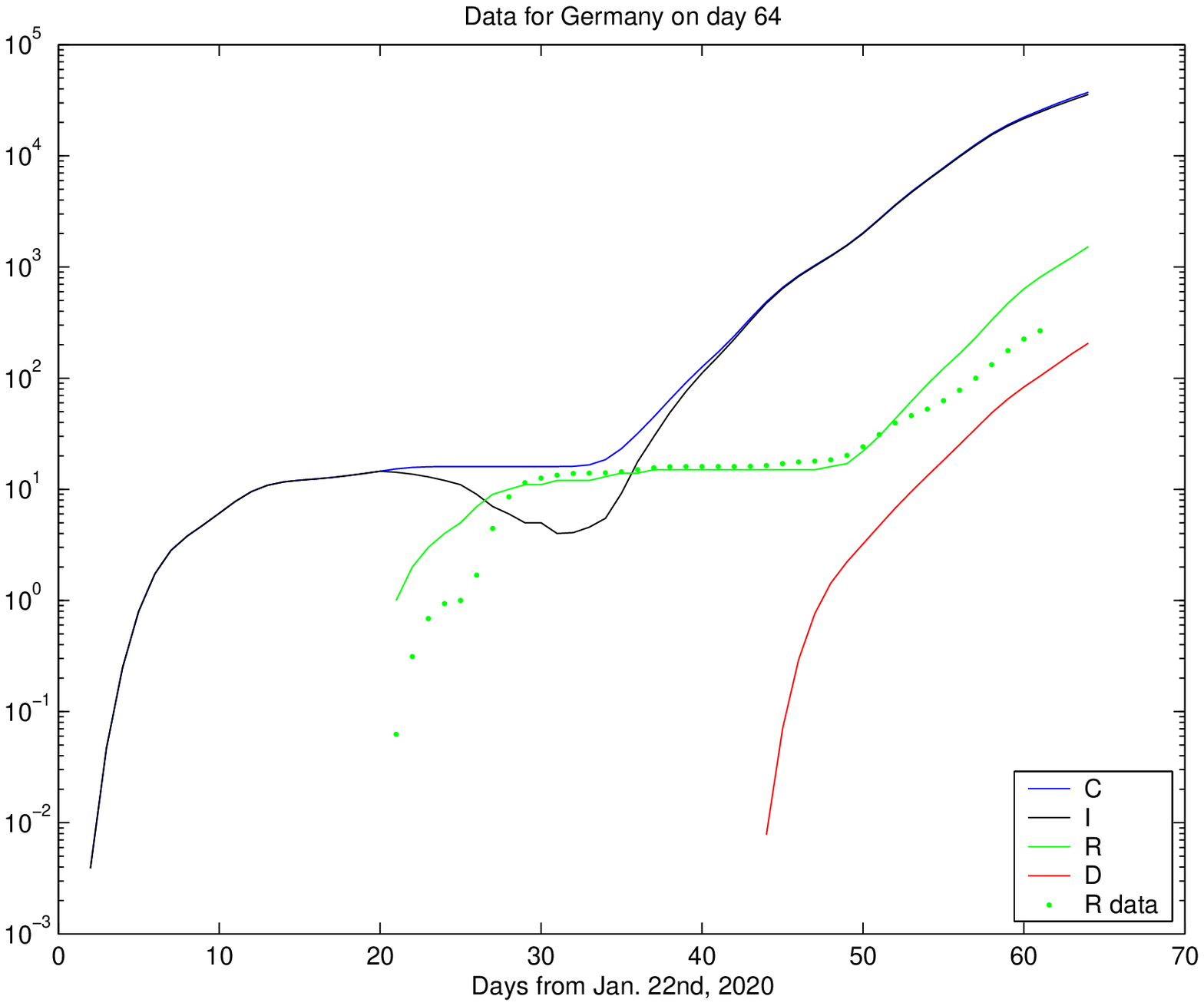}
\caption{Results for France and Germany}
\end{center}  
\end{figure} 
\begin{figure}
\begin{center}
\includegraphics[width=6.0cm,height=5.0cm]{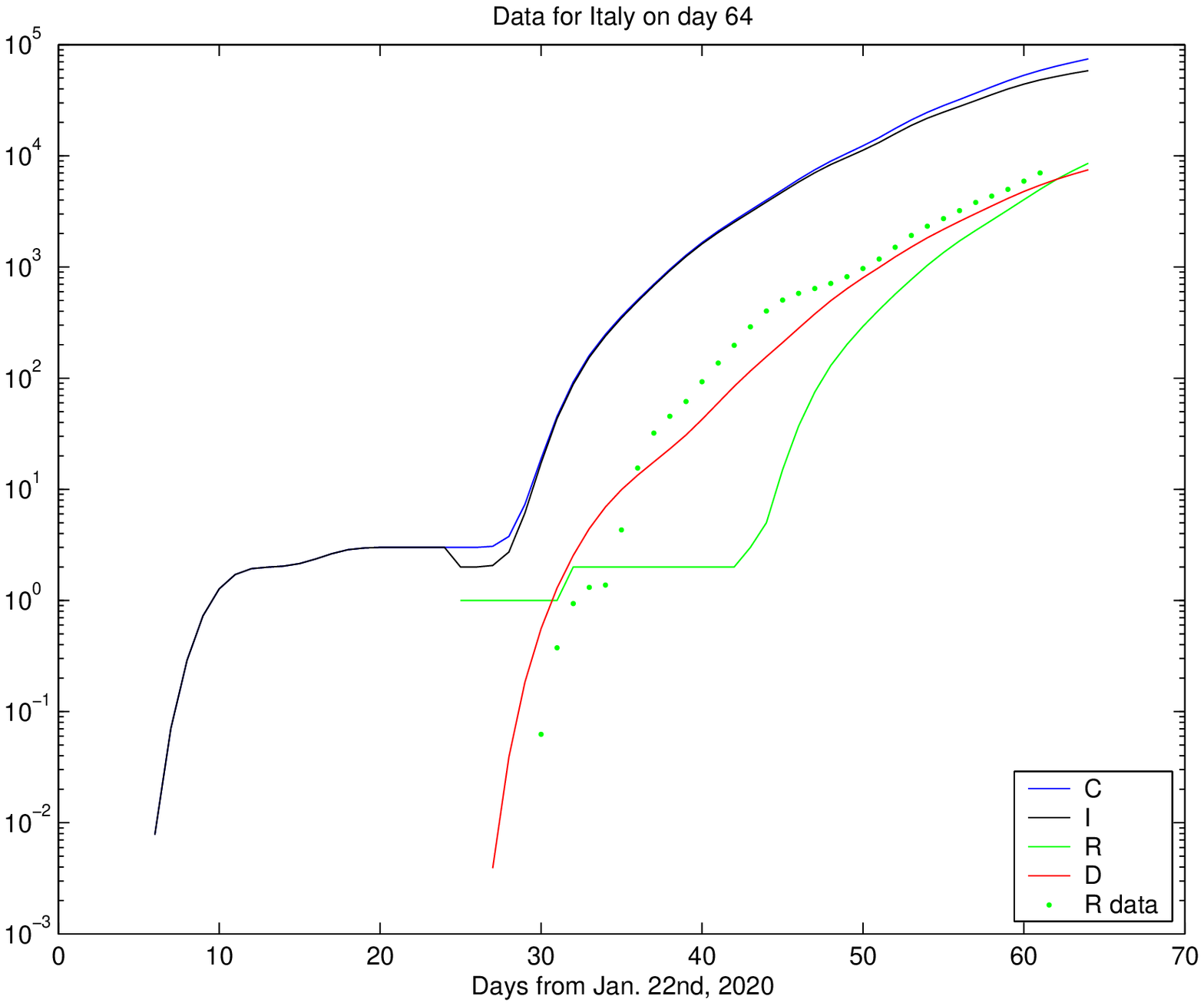}
\includegraphics[width=6.0cm,height=5.0cm]{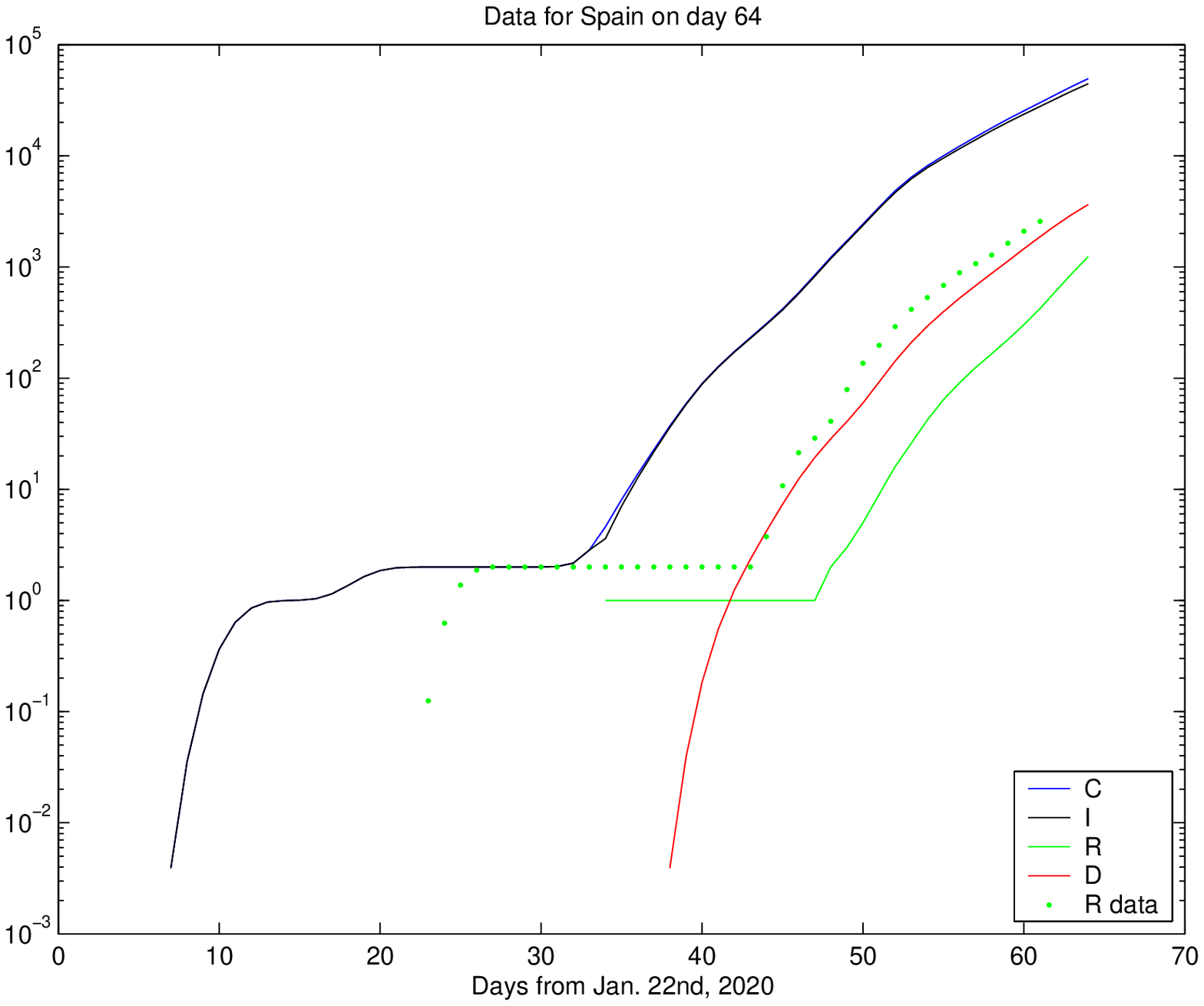}
\caption{Results for Italy and Spain}
\end{center}  
\end{figure} 
\begin{figure}
\begin{center}
\includegraphics[width=6.0cm,height=5.0cm]{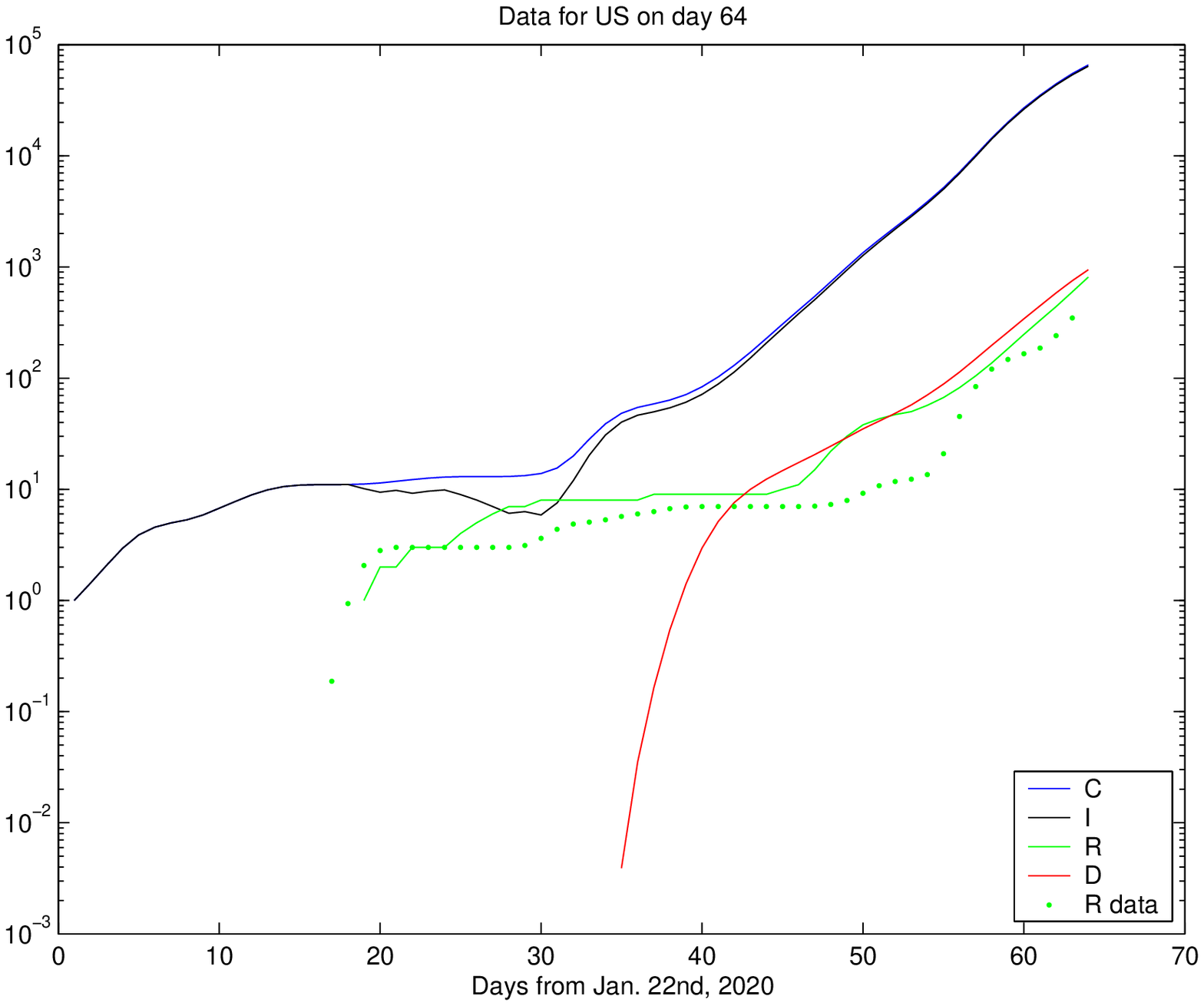}
\includegraphics[width=6.0cm,height=5.0cm]{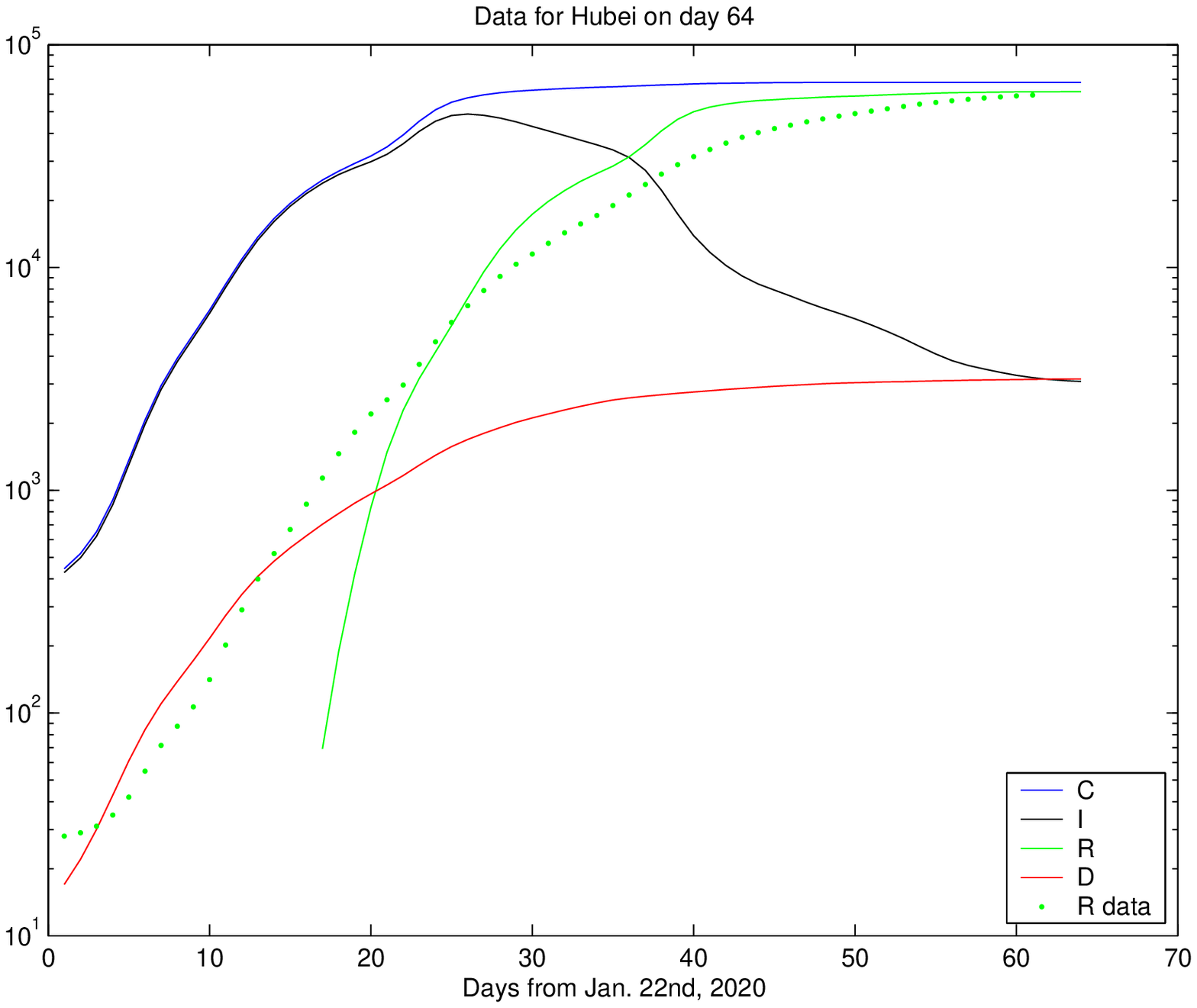}
\caption{Results for US and Hubei}
\end{center}  
\end{figure} 
\bibliographystyle{plain}

\end{document}